\documentclass[english]{achemso}
\usepackage[T1]{fontenc}
\usepackage[utf8]{inputenc}
\usepackage{amsthm}
\usepackage{amsmath}

\makeatletter

\title{Resolution structure in HornSAT and CNFSAT}
\author{KOBAYASHI, Koji}

\numberwithin{equation}{section}
\numberwithin{figure}{section}
\theoremstyle{plain}
\newtheorem{thm}{\protect\theoremname}
  \theoremstyle{definition}
  \newtheorem{defn}[thm]{\protect\definitionname}

\makeatother

\usepackage{babel}
  \providecommand{\definitionname}{Definition}
\providecommand{\theoremname}{Theorem}

\begin{document}

\section{Overview}

This article describes about the difference of resolution structure
and size between HornSAT and CNFSAT.

We can compute HornSAT by using clauses causality. Therefore we can
compute proof diagram by using Log space reduction. But we must compute
CNFSAT by using clauses correlation. Therefore we cannot compute proof
diagram by using Log space reduction, and reduction of CNFSAT is not
P-Complete.

\section{Preparation}

In this paper, we use CNF description as follows;
\begin{defn}
\label{def: Clauses} About $F\in CNF$, we describe the composition
of the clauses $c\in F$ as a subscript. That is, $c_{i\cdots\overline{j}\cdots}=\left(x_{i}\vee\cdots\overline{x_{j}}\vee\cdots\right)$.
The subscript of a capital letter shall be either positive or negative
of a variable. For examples, $c_{I},c_{\overline{I}}$ means $c_{I},c_{\overline{I}}\in\left\{ c_{i},\overline{c_{i}}\right\} ,c_{I}\neq c_{\overline{I}}$
\end{defn}
And define resolution of clauses as follows;
\begin{defn}
\label{def: Resolution} About resolution, I will use the term ``Joint
Variable'' as variables that positive and negative variable which
are included in each antecedents and not included in consequent, and
``Positive Antecedent'' as antecedent that have positive joint variable,
``Negative Antecedent'' as antecedent that have negative joint variable.
We treat some resolution that have same joint variable. Such case,
positive antecedent, negative antecedents and consequents become set
of clauses.
\end{defn}

\section{Resolution}

We show the character of the resolution.
\begin{thm}
\label{thm: Resolution uniformity} In $CNF$ resolution, number of
joint variable of each antecedents is one.\end{thm}
\begin{proof}
I prove it using reduction to absurdity. We assume that some resolution
have 0 or over 2 joint variable. 

The case that resolution have 0 joint variable contradicts a condition
of the resolution clearly. The case that resolution have 2 joint variable
contradicts a condition of the resolution because $c_{IJp\cdots}\vee c_{\overline{I}\overline{J}q\cdots}\rightarrow c_{Jp\cdots\overline{J}q\cdots}=\top$.

Therefore, this theorem was shown than reduction to absurdity.
\end{proof}
I introduce topology of deduction system to formula. For simplification,
I treat topology as formula.
\begin{defn}
\label{def: RCNF} About $F\in CNF$, I will use the term ``DCNF(Deduction
CNF)'' as formula that variables value are presence of restrictions
of CNF formula clauses. Especially, I will use the term ``RCNF(Resolution
CNF)'' and ``$RCNF\left(F\right)$'' as DCNF that deduction system
is resolution principle. Clauses become variables and resolution become
clauses in $RCNF\left(F\right)$. Antecedent become negative variables
and consequent become positive variables. And furthermore, RCNF does
not include variable that correspond to empty clause.
\end{defn}
That is, if

$F\supset c_{ip\cdots}\wedge c_{\overline{i}q\cdots}$,

then

$RCNF\left(F\right)\supset\left(c_{ip\cdots}\right)\wedge\left(c_{\overline{i}q\cdots}\right)\wedge\left(\overline{c_{ip\cdots}}\vee\overline{c_{\overline{i}q\cdots}}\vee c_{p\cdots q\cdots}\right)$.

And RCNF does not include variable correspond to empty clause, therefore
sufficiency of $F$ accords with $RCNF\left(F\right)$. Resolution
consequent is 1 or less, therefore $RCNF\left(F\right)\in HornCNF$.
That is, if $RCNF\left(g\right)=f$;

$RCNF=HornCNF\ni f:\left\{ g\mid g\in CNF\right\} \xrightarrow{Resolution}\left\{ \top,\bot\right\} $

\section{HornSAT and RCNF}

Think $RCNF\left(HornCNF\right)$ complexity. Relation of $HornCNF$
clauses are causality and we can compute them by using unit resolution.
Therefore, we can reduce $HornCNF$ to $RCNF\left(HornCNF\right)$
by using log space reduction. And $RCNF\subset HornCNF$, then $RCNF$
is P-Complete.
\begin{thm}
\label{thm: P-Completeness of RCNF} $f\in P-Complete\mid RCNF\ni f:\left\{ g\mid g\in HornCNF\right\} \xrightarrow{Resolution}\left\{ \top,\bot\right\} $\end{thm}
\begin{proof}
Clearly $RCNF\subset HornCNF$ and $RCNF\in P$, I should show that
$\exists h\in L\left(h:g\mapsto f\right)$ (L:Log space reduction).
We treat $h$ as 2-step procedures to simplify this. First, 

First, I reduce $HornCNF$ to at most 3 variables clauses $HornCNF$.
We can reduce by using same way to reduce $CNF$ to $3CNF$. That
is, each clauses change follows with new variables.

$g\ni c_{I\overline{j}\overline{k}\overline{l}\cdots}\rightarrow c_{I\overline{j}\overline{0}}\wedge c_{0\overline{k}\overline{1}}\wedge c_{1\overline{l}\overline{2}}\wedge\cdots\in g'$

We can execute this reduction with logarithm space, pointer to consequent,
pointer to variable, counter that show already used variables.

Second, I reduce $c'\in g'$ to $RCNF\left(c'\right)$. We can reduce
by adding resolution formula for each clauses. We can reduce HornCNF
with unit resolution, therefore it is enough to keep SAT by using
resolution formula that variables of antecedent decreases. That is;

$c_{R}\rightarrow\left(x_{R}\right)\wedge\left(\overline{x_{R}}\vee\overline{x_{\overline{R}}}\right)$

$c_{P\overline{q}}\rightarrow\left(x_{P\overline{q}}\right)\wedge\left(x_{P}\vee\overline{x_{P\overline{q}}}\vee\overline{x_{q}}\right)\wedge\left(\overline{x_{P}}\vee\overline{x_{\overline{P}}}\right)$

$c_{I\overline{j}\overline{k}}\rightarrow\left(x_{I\overline{j}\overline{k}}\right)\wedge\left(x_{I\overline{k}}\vee\overline{x_{I\overline{j}\overline{k}}}\vee\overline{x_{j}}\right)\wedge\left(x_{I\overline{j}}\vee\overline{x_{I\overline{j}\overline{k}}}\vee\overline{x_{k}}\right)\wedge\left(x_{I}\vee\overline{x_{I\overline{j}}}\vee\overline{x_{j}}\right)\wedge\left(x_{I}\vee\overline{x_{I\overline{k}}}\vee\overline{x_{k}}\right)\wedge\left(\overline{x_{I}}\vee\overline{x_{\overline{I}}}\right)$

We can execute this reduction with logarithm space, pointer to consequent,
pointer to variable.

Above two reduction, we can reduce $HornCNF$ to $RCNF$. Both reductions
use only logarithm space, we can execute all reduction $h:g\mapsto g'\mapsto f$
in logarithm space.

Therefor, RCNF is P-Complete.
\end{proof}

\section{CNFSAT and RCNF}

Think $RCNF\left(CNF\right)$ complexity. Relation of $CNF$ clauses
are correlation and we cannot compute them by using unit resolution.
Therefore, we cannot reduce $CNF$ to $RCNF\left(CNF\right)$ by using
log space reduction. And $RCNF\subset HornCNF$. That is, $RCNF$
is not P-Complete.

Afterward, we show some $CNF$ that $RCNF$ is not P-Complete. First,
we think the formula that each reduction depend whole formula.
\begin{defn}
\label{def: SCNF} We will use the term ``$S3CNF$(3-Simplex CNF)''
to;

$t_{PQR}=c_{\overline{P}\overline{Q}}\wedge c_{\overline{Q}\overline{R}}\wedge c_{\overline{P}\overline{R}}\wedge c_{PQR}$

and ``$S4CNF$(4-Simplex CNF)'' to;

$T_{PQR}=c_{\overline{P}\overline{Q}R}\wedge c_{P\overline{Q}\overline{R}}\wedge c_{\overline{P}Q\overline{R}}\wedge c_{PQR}$

and ``$SCNF$'' to $S3CNF\cup S4CNF$.
\end{defn}
Second, we think the formula that consist of $SCNF$.
\begin{defn}
\label{def: CCNF} $f\in CNF$ that consist of $SCNF$, we will use
term ``$CCNF$(Chaotic CNF)'' if $f$ satisfy follow condition.

The Graph that each $SCNF\ni t\subset f$ are nodes and each variables
are edges.

a) This Graph is 3-Moore Graph.

b) When this graph girth is $2k+1$, all circuit include $S4CNF$
that number is $k\times c_{0}\mid c_{0}:const\left(c_{0}>1\right)$.
\end{defn}
Next, we think that $RCNF\left(CCNF\right)$ is not P-Complete. We
show that $RCNF\left(CCNF\right)$ is not polynomial size and we cannot
treat $RCNF\left(CCNF\right)$ by using log space reduction. 
\begin{thm}
\label{thm: CCNF size} $f\in CCNF$ are exists that $RCNF\left(CCNF\right)$
is not polynomial size.\end{thm}
\begin{proof}
I prove it using reduction to absurdity. We assume that we can reduce
all all $f\in CCNF$ to $RCNF\left(CCNF\right)$ in polynomial size.
From this assumption, number of $RCNF\left(CCNF\right)$ consequent
stay in polynomial size.

From $S4CNF$ structure, each $S4CNF$ resolution's consequents are
include over one joint variables. Therefore, next resolution must
include another clause as antecedent. That is, $S4CNF$ resolution
become product of positive antecedents and negative antecedents. And
$f$ is Moore Graph structure, therefore it is necessary over girth
$2k+1$ clauses to appear same clause in processing resolution antecedent.
Resolution that one of antecedent is $S4CNF$ have consequents size
twice of antecedents size. Therefore, consequents size become $2^{k\times c_{0}}$.
On the other side, size of 3-Moore Graph is $1+3\overset{k-1}{\underset{i=0}{\sum}}(3-1)^{i}=1+3\times\left(2^{k}-1\right)$.
Therefore, ratio of size of $f$ and consequents of $RCNF\left(f\right)$
is;

$O\left(\dfrac{\left|f\right|}{\left|RCNF\left(f\right)\right|}\right)=O\left(\dfrac{2^{k\times c_{0}}}{1+3\times\left(2^{k}-1\right)}\right)\rightarrow O\left(c^{k}\right)\quad\left(as\quad k\gg0\right)$

And $RCNF\left(f\right)$ consequents is not in polynomial size and
contradicts a condition that $RCNF\left(CCNF\right)$ consequent stay
in polynomial size.

Therefore, this theorem was shown than reduction to absurdity.\end{proof}
\begin{thm}
\label{thm: L limit}$\left(RCNF\ni f:\left\{ g\mid g\in CCNF\right\} \xrightarrow{Resolution}\left\{ \top,\bot\right\} \right)\rightarrow\left(\forall h\in L\left(h:g\not\mapsto f\right)\right)$\end{thm}
\begin{proof}
I prove it using reduction to absurdity. We assume that $h$ exists
that all $g\mapsto f$ satisfy this theorem. Because $h\in L$, $h$
classify at most polynomial size. Therefore, size of $f$ (that is
target of $h$) also stay polynomial size.

But mentioned above \ref{thm: CCNF size}, $CCNF$ have $f$ that
is not in polynomial size. Therefore, there exists $f$ that is $L\not\ni h:g\longrightarrow f$
and contradicts a condition that $h$ exist that all $f$ of $g\mapsto f$
in polynomial size.

Therefore, this theorem was shown than reduction to absurdity.\end{proof}
\begin{thm}
\label{thm: RCNF limit}$f\not\in P-Complete\mid RCNF\ni f:\left\{ g\mid g\in CCNF\right\} \xrightarrow{Resolution}\left\{ \top,\bot\right\} $\end{thm}
\begin{proof}
Mentioned above \ref{thm: L limit}, there is no log space reduction
that reduce $g\in CCNF$ to $f\in RCNF$. Therefore, $f$ is not P-Complete.\end{proof}

\end{document}